\documentclass[journal]{new-aiaa}
\usepackage[utf8]{inputenc}
\usepackage{bm}
\usepackage{graphicx}
\usepackage{amsmath}
\usepackage[version=4]{mhchem}
\usepackage{siunitx}
\usepackage{float}
\usepackage{multirow}
\usepackage{longtable,tabularx}
\usepackage{setspace}
\title{Reflective-Sail Weak Stability Boundary Structure with the Locally Optimal Control Law}

\author{Shuyue Fu \footnote{PhD Candidate, School of Astronautics, fushuyue@buaa.edu.cn.}, Jinkai Zhang \footnote{Graduate Student, School of Astronautics, sy2515109zjk@buaa.edu.cn.}, Di Wu \footnote{Associate Professor, School of Astronautics, wudi2025@buaa.edu.cn, Member AIAA.}, Shengping Gong \footnote{Professor, School of Astronautics, gongsp@buaa.edu.cn, Senior Member AIAA (Corresponding Author).}, and Peng Shi\footnote{Professor, School of Astronautics, shipeng@buaa.edu.cn.}}
\affil{Beihang University, Beijing, 100191, People's Republic of China}
\affil{State Key Laboratory of High-Efficiency Reusable Aerospace Transportation Technology, Beijing, 102206, People's Republic of China}

\begin{document}

\maketitle

\begin{abstract}
Escaping from the Earth is the first step of interplanetary transfers. Traditional ballistic escape trajectories in the Sun-Earth circular restricted three-body problem face limitations in relatively long time of flight and low hyperbolic excess velocity. To augment the construction of escape trajectories from the Earth, this Note proposes the concept of reflective-sail weak stability boundary structures and accordingly constructs and analyzes escape trajectories from the Earth in the context of the Sun-Earth planar circular restricted three-body problem with a reflective sail. Using an ideal reflective sail, the locally optimal control law to maximize the time derivative of the Keplerian energy with respect to the Earth is adopted. Levi-Civita regularization about the Earth is derived to address the singularity caused by the Earth. The configurations of reflective-sail weak stability boundary structures are calculated to provide initial states for constructing escape trajectories and information about regions where escape is facilitated. Then, the escape trajectories using a reflective sail are constructed based on the proposed weak stability boundary structures. The escape performance, including time of flight and estimated hyperbolic excess velocity, is analyzed. Comparison with ballistic escape trajectories in the Sun-Earth PCR3BP is also performed, indicating improved escape performance characterized by shorter time of flight and higher hyperbolic excess velocity.
\end{abstract}

\section{Introduction}
\lettrine{E}{scape} from the Earth is the first step to explore deep space. Constructing escape trajectories achieving a trade-off between efficiency and fuel consumption is an important topic in trajectory design. To construct escape trajectories with low fuel consumption, the natural force, including multi-body gravity \cite{topputo2008resonant,casalino2020design,oshima2021capture,fu2026energy} and solar radiation pressure (SRP) \cite{mengali2004earth,macdonald2005realistic,terzaghi2026performance}, should be sufficiently considered and utilized in the escape trajectory construction. When considering the multi-body gravity, a typical type of escape trajectory is the ballistic escape trajectory \cite{topputo2008resonant}, which is initially at the temporary capture state with respect to a planet, and then escapes to the heliocentric space without maneuver. 
Compared with the purely multi-body gravity, the use of solar sail spacecraft provides an additional SRP acceleration, enabling a more efficient escape than a ballistic escape trajectory. Solar sail spacecraft can be categorized into reflective sail \cite{mcinnes2004solar,macdonald2005realistic,macdonald2005analytical}, refractive sail \cite{firuzi2021gradient,wang2025optimal}, and diffractive sail \cite{chu2024minimum,terzaghi2026performance}. Among such types of solar sail spacecraft, the technology of reflective sail is the most mature and has been applied to the practical mission (e.g., IKAROS \cite{mori2014overview}). Moreover, although refractive sail and diffractive sail can effectively utilize tangential SRP acceleration to change the semi-major axis of the heliocentric orbit, the attitude of solar sail spacecraft should keep Sun-facing according to the model of these sails \cite{firuzi2021gradient,chu2024minimum}. For the Sun-facing attitude, the tangential SRP acceleration may not be effectively utilized in the planetary escape scenario. This scenario usually effectively changes the semi-major axis of the orbit/trajectory around the planet, rather than around the Sun. In contrast, the reflective sail allows for a more flexible attitude maneuver to effectively utilize the SRP acceleration to make the spacecraft escape from the Earth. For example, Macdonald and McInnes \cite{macdonald2005analytical} developed the locally optimal control laws to effectively use the SRP acceleration, including the locally optimal control law to achieve the escape from the Earth. For these reasons, to achieve efficient escape from the Earth, this Note considers the construction problem of escape trajectories from the Earth using a reflective sail. Comparing the magnitude of the SRP acceleration and the acceleration of the third-body perturbation from the Sun, it is found that the magnitude of the third-body perturbation acceleration from the Sun is larger \cite{mora2020solar}. Therefore, the escape trajectories can be constructed in the context of the Sun-Earth circular restricted three-body problem (CR3BP) with a reflective sail.

Several studies have investigated the CR3BP (also other multi-body problems, such as the elliptical restricted three-body problem) with a reflective sail. Researchers typically focused on the use of a reflective sail in the generation of artificial libration points \cite{mcinnes1994solar,mcinnes2004solar,baoyin2006solar}, periodic or quasi-periodic orbits \cite{mora2020solar,chujo2024quasi,hettrick2025approximate}, and invariant manifolds and corresponding transfers \cite{waters2008invariant,gong2010solar,hettrick2025approximate}. For the escape trajectory construction, Coverstone and Prussing \cite{coverstone2003technique} derived the locally optimal control law to maximize the time derivative of orbital energy, and constructed escape trajectories from the geosynchronous transfer orbit. Mengali and Quarta \cite{mengali2004earth} took more perturbations than the third-body perturbation from the Sun into account, and Macdonald and McInnes \cite{macdonald2005realistic} proposed realistic Earth escape strategies, including the high-fidelity dynamical model, Earth shadow, and blended locally optimal control laws. The aforementioned works \cite{coverstone2003technique,mengali2004earth,macdonald2005realistic} generally consider the third-body perturbation from the Sun. Therefore, these works can essentially be considered as multi-body trajectory construction problems using a reflective sail. However, the solution space of escape trajectories in the Sun-Earth PCR3BP with a reflective sail, including the time of flight (TOF) and hyperbolic excess velocity, has not been analyzed systematically. Moreover, these works addressed the construction problem of escape trajectories from the specific Earth orbits (e.g., geosynchronous transfer orbit and geosynchronous Earth orbit) and have not provided the information about which region can facilitate escape or can not facilitate escape compared with the multi-body perturbations. To bridge this gap, this Note is devoted to developing a framework to perform a systematic analysis of escape trajectories using a reflective sail in the multi-body dynamics. In the context of the purely multi-body dynamics, the analysis of capture/escape trajectories can be performed using the invariant manifolds \cite{Koon2001,qian2016energy} or weak stability boundary (WSB) structures \cite{belbruno1993sun,hyeraci2010method}. Compared with invariant manifolds in the PCR3BP, the WSB structure can be extended to other dynamical models without any limitation \cite{hyeraci2010method}. Meanwhile, when calculating WSB structures, the use of osculating eccentricity with respect to the secondary body (e.g., the Earth in this Note) can further make matching mission constraints more convenient \cite{hyeraci2010method}. Based on the aforementioned discussion, the WSB structure can provide a useful framework for understanding the dynamical behaviors around the Earth and mission design without the requirement of prior knowledge about the periodic orbits and their invariant manifolds. This motivates us to propose the concept of reflective-sail WSB structures and to develop an analysis framework for escape trajectories considering both the SRP acceleration and multi-body gravity.

In this Note, we adopt the Sun-Earth planar CR3BP (PCR3BP) with a reflective sail as the dynamical model. Levi-Civita regularization about the Earth is derived to address the singularity caused by the Earth, which can be considered as a contribution to the modeling of the reflective sail multi-body problem. Then, the concept of reflective-sail WSB structures is proposed, and the configurations of reflective-sail WSB structures are obtained. To effectively utilize the SRP acceleration during the escape, we adopt the locally optimal control law to maximize the time derivative of the Keplerian energy with respect to the Earth \cite{coverstone2003technique,macdonald2005analytical}. The configurations of reflective-sail WSB structures provide information about the potential initial state distribution of escape trajectories. A comparison with the corresponding regions in the Sun–Earth PCR3BP identifies regions where a reflective sail facilitates escape and regions where it does not. An interesting finding is that the use of a reflective sail may not always facilitate escape from the Earth, even with the locally optimal control law. Subsequently, following the method proposed by Hyeraci and Topputo \cite{hyeraci2010method}, the construction method of escape trajectories is proposed by the intersection of the backward stable sets in the Sun-Earth PCR3BP and the forward unstable sets using a reflective sail. The corresponding escape performance (TOF and estimated hyperbolic excess velocity) is analyzed to reveal the advantages of the introduction of a reflective sail with the locally optimal control law. Pareto fronts of the obtained solutions in terms of TOF and estimated hyperbolic excess velocity are extracted. Results show that compared with ballistic escape trajectories in the Sun-Earth PCR3BP, the escape trajectories using a reflective sail generally have better escape performance as Pareto fronts using a reflective sail are trended to be distributed in the upper-left corner of those in the Sun-Earth PCR3BP, indicating shorter TOF and higher hyperbolic excess velocity. The typical escape trajectory is finally analyzed.

The rest of this Note is organized as follows. Section \ref{sec2} presents the dynamical model of this work. Section \ref{sec3} introduces the concept of the reflective-sail WSB structure and investigates its configurations. Section \ref{sec4} proposes the method to construct escape trajectories based on the reflective-sail WSB structures. The corresponding escape performance is analyzed. Finally, conclusions are drawn in Section \ref{sec5}.

\section{Sun-Earth PCR3BP with a Reflective Sail}\label{sec2}
This section presents the dynamical model of the Sun-Earth PCR3BP with a reflective sail, including the dynamical equations, Levi-Civita regularization, and the locally optimal control law of pitch angle.
\subsection{Dynamical Model}\label{sec2.1}
\begin{figure}[H]
\centerline{\includegraphics[width=0.3\textwidth]{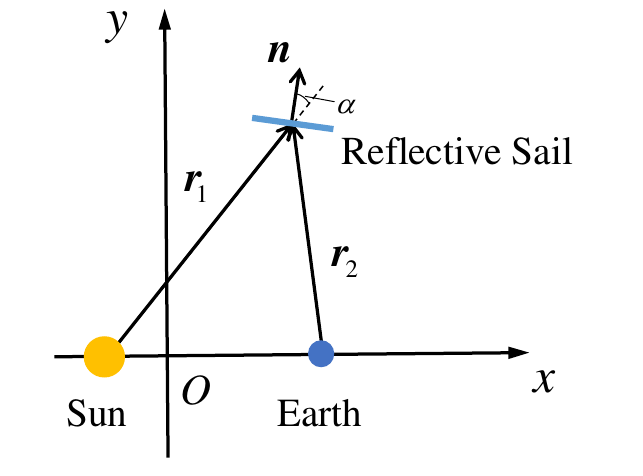}}
\caption{Schematic of the Sun-Earth PCR3BP with a reflective sail.}\label{fig_PCR3BP}
\end{figure}

We adopt the Sun-Earth PCR3BP with a reflective sail to investigate the reflective-sail WSB structures. The Sun-Earth rotating frame is adopted (as shown in Fig. \ref{fig_PCR3BP}), and the dimensionless units are set as follows: the mass unit (MU) is the combined mass of the Sun and the Earth, the length unit (LU) is the Sun-Earth distance, and the time units (TU) is set as $T_{\text{SE}}/2\pi$ ($T_{\text{SE}}$ denotes the orbital period of the Sun/Earth around the Sun-Earth barycenter).

Then, the dynamical equations can be expressed as:
\begin{equation}
  \dot x = u ,\text{ }\dot y = v,\text{ }\dot u = x + 2v - \frac{{\left( {1 - \mu } \right)\left( {x + \mu } \right)}}{{{r_1}^3}} - \frac{{\mu \left( {x + \mu  - 1} \right)}}{{{r_2}^3}} + {a_{{\text{SRP}}x}},\text{ }\dot v = y - 2u - \frac{{\left( {1 - \mu } \right)y}}{{{r_1}^3}} - \frac{{\mu y}}{{{r_2}^3}} + {a_{{\text{SRP}}y}}\label{eq1}
\end{equation}
\begin{equation}
{r_1} = \sqrt {{{\left( {x + \mu } \right)}^2} + {y^2}} \text{ }\text{ }\text{ }\text{ }{r_2} = \sqrt {{{\left( {x + \mu - 1} \right)}^2} + {y^2}}\label{eq2}
\end{equation}
where $\mu$ denotes the mass parameter of the Sun-Earth system, $\bm{X}=\left[x,\text{ }y,\text{ }u,\text{ }v\right]^{\text{T}}$ denotes the state vector, $r_1$ and $r_2$ denote the distances between the reflective sail and the Sun/Earth, and $a_{\text{SRP}}$ denotes the SRP acceleration generated by a reflective sail. According to Ref. \cite{chu2024potential}, the SRP acceleration generated by a reflective sail (ideal reflective sail) for the planar case can be expressed as:
\begin{equation}
  {a_{{\text{SRP}}x}} =  \frac{{\beta \left( {1 - \mu } \right)}}{{{r_1}^3}}{\cos ^2}\alpha\left( {\left( {x + \mu } \right)\cos \alpha  - y\sin \alpha } \right) ,\text{ }
  {a_{{\text{SRP}}y}} =  \frac{{\beta \left( {1 - \mu } \right)}}{{{r_1}^3}}{\cos ^2}\alpha\left( {y\cos \alpha  + \left( {x + \mu } \right)\sin \alpha } \right) 
\label{eq3}
\end{equation}
where $\beta$ denotes the sail’s lightness number (whose value is usually up to 0.05 for contemporary highest estimations \cite{chu2024potential}), and $\alpha$ denotes the pitch angle defined as the angle between the direction of sunlight $\bm{r}_1$ and the sail’s normal direction $\bm{n}$ ($\alpha \in \left[-\frac{\pi}{2},\text{ }\frac{\pi}{2}\right]$). In this model, the Jacobi energy is time-varying, which can be expressed as:
\begin{equation}
C = -\left( {{u^2} + {v^2}} \right) +  \left( {{x^2} + {y^2}} \right) + \frac{{2(1 - \mu) }}{{{r_1}}} + \frac{2\mu }{{{r_2}}} + \mu \left( {1 - \mu } \right) \label{eq4}
\end{equation}
When investigating the dynamical behaviors around the Earth, the cases where trajectories approach the Earth should be addressed. When $r_2 \to 0$, the dynamical equations shown in Eq. \eqref{eq1} become singular because the term $\frac{1}{r_2^3}$. To address the singularity and accordingly improve the computation efficiency and accuracy, Levi-Civita regularization about the Earth is derived. The regularized variables are adopted as follows \cite{oshima2017analysis}:
\begin{equation}
  x - 1 + \mu  = {u_1}^2 - {u_2}^2 ,\text{ }y = 2{u_1}{u_2},\text{ }u = \frac{{2\left( {{u_1}{u_3} - {u_2}{u_4}} \right)}}{{{u_1}^2 + {u_2}^2}},\text{ } v = \frac{{2\left( {{u_2}{u_3} + {u_1}{u_4}} \right)}}{{{u_1}^2 + {u_2}^2}},\text{ }{\text{d}}t = {r_2}{\text{d}}s 
\label{eq5}
\end{equation}
With these variables, the regularized dynamical equations can be expressed as:
\begin{equation}
  {u_1}^\prime  = {u_3} ,\text{ }{u_2}^\prime  = {u_4},\text{ }{u_3}^\prime  = \frac{1}{4}\left( {a + b} \right){u_1} + \frac{1}{4}c{u_2} + {\Delta _1},\text{ }{u_4}^\prime  = \frac{1}{4}\left( {a - b} \right){u_2} + \frac{1}{4}c{u_1} + {\Delta _2}
\label{eq6}
\end{equation}
where:
\begin{equation}\label{eq7}
a = \frac{{2\left( {1 - \mu } \right)}}{{\sqrt {{{\left( {{u_1}^2 - {u_2}^2 + 1} \right)}^2} + 4{u_1}^2{u_2}^2} }} - C + {\left( {{u_1}^2 - {u_2}^2 + 1 - \mu } \right)^2} + 4{u_1}^2{u_2}^2 + \mu \left( {1 - \mu } \right)
\end{equation}
\begin{equation}
b = 8\left( {{u_2}{u_3} + {u_1}{u_4}} \right)+ 2\left( {{u_1}^2 + {u_2}^2} \right)\left( {{u_1}^2 - {u_2}^2 + 1 - \mu } \right) - \frac{{2\left( {1 - \mu } \right)\left( {{u_1}^2 - {u_2}^2 + 1} \right)\left( {{u_1}^2 + {u_2}^2} \right)}}{{{{\left( {\sqrt {{{\left( {{u_1}^2 - {u_2}^2 + 1} \right)}^2} + 4{u_1}^2{u_2}^2} } \right)}^3}}}
\label{eq8}
\end{equation}
\begin{equation}\label{eq9}
c = 4{u_1}{u_2}\left( {{u_1}^2 + {u_2}^2} \right) - 8\left( {{u_1}{u_3} - {u_2}{u_4}} \right)  - \frac{{4\left( {1 - \mu } \right){u_1}{u_2}\left( {{u_1}^2 + {u_2}^2} \right)}}{{{{\left( {\sqrt {{{\left( {{u_1}^2 - {u_2}^2 + 1} \right)}^2} + 4{u_1}^2{u_2}^2} } \right)}^3}}}
\end{equation}
\begin{equation}\label{eq10}
{\Delta _1} = \frac{{\beta  \left( {1 - \mu } \right){{\cos }^2}\alpha\left( {{u_1}^2 + {u_2}^2} \right)}}{{2{{\left( {\sqrt {{{\left( {{u_1}^2 - {u_2}^2 + 1} \right)}^2} + 4{u_1}^2{u_2}^2} } \right)}^3}}}\left( {{u_1}\left( {1 + {u_1}^2 + {u_2}^2} \right)\cos \alpha  + {u_2}\left( {1 - {u_1}^2 - {u_2}^2} \right)\sin \alpha } \right)
\end{equation}
\begin{equation}\label{eq11}
{\Delta _2} = \frac{{\beta  \left( {1 - \mu } \right){{\cos }^2}\alpha\left( {{u_1}^2 + {u_2}^2} \right)}}{{2{{\left( {\sqrt {{{\left( {{u_1}^2 - {u_2}^2 + 1} \right)}^2} + 4{u_1}^2{u_2}^2} } \right)}^3}}}\left( {{u_2}\left( {{u_1}^2 + {u_2}^2 - 1} \right)\cos \alpha  + {u_1}\left( {1 + {u_1}^2 + {u_2}^2} \right)\sin \alpha } \right)
\end{equation}
In Eq. \eqref{eq6}, $\left(\cdot\right)^\prime$ denotes $\text{d}\left(\cdot\right)/\text{d}s$. To solve Eq. \eqref{eq6}, the derivative of $C$ should be combined:
\begin{equation}\label{eq12}
C' = \frac{{ - 4\beta  \left( {1 - \mu } \right){{\cos }^2}\alpha}}{{{{\left( {\sqrt {{{\left( {{u_1}^2 - {u_2}^2 + 1} \right)}^2} + 4{u_1}^2{u_2}^2} } \right)}^3}}}\left( \begin{gathered}
  {u_3}\left( {{u_1}\left( {1 + {u_1}^2 + {u_2}^2} \right)\cos \alpha  + {u_2}\left( {1 - {u_1}^2 - {u_2}^2} \right)\sin \alpha } \right) \hfill \\
   + {u_4}\left( {{u_2}\left( {{u_1}^2 + {u_2}^2 - 1} \right)\cos \alpha  + {u_1}\left( {1 + {u_1}^2 + {u_2}^2} \right)\sin \alpha } \right) \hfill \\ 
\end{gathered}  \right)
\end{equation}
For numerical integration, we adopt a variable-step, variable-order (VSVO) Adams-Bashforth-Moulton predictor–corrector method (orders 1 to 13) to integrate trajectories. The absolute and relative tolerances are set to $1\times 10^{-13}$. Parameters used in the simulations are summarized in Table \ref{tab_parameter} \footnote{$\text{https://ssd.jpl.nasa.gov/tools/periodic\_orbits.html}$}.
\begin{table}[!htb]
\caption{ Parameters used in the simulations}\label{tab_parameter}%
\centering
\renewcommand{\arraystretch}{1.5}
\begin{tabular}{@{}llll@{}}
\hline
Symbol & Value  & Units & Meaning\\
\hline
$\mu$    & $3.0542 \times {10^{ - 6}}$   & --  & Sun-Earth mass parameter  \\
$R_{\text{E}}$    & $6378$   & km  & Earth’s radius  \\
LU    & $1.49597871 \times {10^8}$   & km  & Length unit  \\
TU    & $5.022635 \times {10^6}$   & s  & Time unit  \\
\hline
\end{tabular}
\end{table}

\subsection{Locally Optimal Control Law of Pitch Angle}\label{sec2.2}
To achieve efficient escape from the Earth, the locally optimal control law of pitch angle is adopted. The locally optimal control law ensures the maximum variation of the two-body orbital elements or the corresponding quantities with respect to the planet when the state $\bm{X}$ is presented \cite{macdonald2005analytical}. In this Note, we investigate the locally optimal control law to maximize the time derivative of the Keplerian energy with respect to the Earth ($E_2$). The Keplerian energy with respect to the Earth can be expressed as:
\begin{equation}
{E_{2}} = \frac{1}{2}\left( {{{\left( {{u} - {y}} \right)}^2} + {{\left( {{v} + {x} + \mu  - 1} \right)}^2}} \right) - \frac{\mu }{r_{2}} \label{eq13}
\end{equation}
In the context of the Sun-Earth PCR3BP with a reflective sail, the time derivative of $E_2$ caused by the reflective sail can be expressed as:
\begin{equation}\label{eq_control_law_2}
{{\dot E}_{{\text{2sail}}}} = \frac{{\beta \left( {1 - \mu } \right)}}{{{r_1}^3}}{\cos ^2}\alpha \left( {A\cos \alpha  + B\sin \alpha } \right)
\end{equation}
\begin{equation}\label{eq_control_law_2_new}
  A = \left( {u - y} \right)\left( {x + \mu } \right) + \left( {v + x + \mu  - 1} \right)y,\text{ }
  B =  - \left( {u - y} \right)y + \left( {v + x + \mu  - 1} \right)\left( {x + \mu } \right) 
\end{equation}
Due to ${E'_{{\text{2sail}}}} = {r_2}{\dot E_{{\text{2sail}}}}$, i.e., $\max{\dot E_{{\text{2sail}}}} \leftrightarrow \max {E'_{{\text{2sail}}}}$ when the state $\bm{X}$ is presented, the following analysis of the locally optimal control law focuses on ${\dot E_{{\text{2sail}}}}$. Introduce $\tilde{\alpha}$:
\begin{equation}\label{eq_control_law_3}
  \tilde \alpha  = {\text{atan2}}\left( {B,{\text{ }}A} \right) ,\text{ } A = \rho \cos \tilde \alpha,\text{ }B = \rho \sin \tilde \alpha
\end{equation}
The problem of obtaining $\max{\dot E_{{\text{2sail}}}}$ when the $\bm{X}$ is presented can be further transformed into the following problem ($\rho >0$):
\begin{equation}\label{eq_control_law_4}
\mathop {\max }\limits_\alpha  {\cos ^2}\alpha \cos \left( {\alpha  - \tilde \alpha } \right)
\end{equation}
Therefore, when $\alpha\ \in \left(-\frac{\pi}{2},\text{ }\frac{\pi}{2}\right)$, the locally optimal control law $\alpha_{\text{LO}}\left(t\right)$ satisfies:
\begin{equation}\label{eq_control_law_5}
\tan \alpha_{\text{LO}} \left( t \right) = \frac{{ - 3\cos \tilde \alpha  + \sqrt {9{{\cos }^2}\tilde \alpha  + 8{{\sin }^2}\tilde \alpha } }}{{4\sin \tilde \alpha }}{\text{ }}\left( {B  \ne 0} \right)
\end{equation}
where the subscript “LO” represents the locally optimal control law. Equation \eqref{eq_control_law_5} is in the same form as the locally optimal control law in Refs. \cite{macdonald2005realistic,macdonald2005analytical}. Notably, when $B$ approaches 0, the control law shown in Eq. \eqref{eq_control_law_5} becomes degenerate. The locally optimal control law for this case is determined by the boundary candidates, i.e., $\alpha_{\text{LO}}=0$ or $\alpha_{\text{LO}}  =  \pm \frac{\pi }{2}$. Above all, the complete locally optimal control law $\alpha_{\text{LO}}\left(t\right)$ can be expressed as:
\begin{equation}\label{eq_control_law_6}
\alpha_{\text{LO}} \left( t \right) = \left\{ {\begin{array}{*{20}{c}}
  {0{\text{ }\text{ }\text{ }}A > 0{\text{ and }}B = 0} \\ 
  { \pm \frac{\pi }{2}{\text{   }\text{ }\text{ }}A < 0{\text{ and }}B = 0} \\ 
  {{\text{atan2}}\left( {\frac{\tau }{{\sqrt {{\tau ^2} + 1} }},{\text{ }}\frac{1}{{\sqrt {{\tau ^2} + 1} }}} \right){\text{   }\text{ }\text{ }}A \ne 0{\text{ and }}B \ne 0} 
\end{array}} \right.
\end{equation}
where:
\begin{equation}\label{eq_control_law_7}
\tau  = \frac{{ - 3\cos \tilde \alpha  + \sqrt {9{{\cos }^2}\tilde \alpha  + 8{{\sin }^2}\tilde \alpha } }}{{4\sin \tilde \alpha }}
\end{equation}
Although $\alpha_{\text{LO}}\left(t\right)$ can exhibit discontinuity, ${a_{{\text{SRP}}x}}$ and ${a_{{\text{SRP}}y}}$ remain continuous during the flight when $\max \left(\left|A\right|,\text{ }\left|B\right|\right)>0$. Therefore, the VSVO Adams-Bashforth-Moulton predictor–corrector method (orders 1 to 13) is also suitable for integrating reflective-sail trajectories with the locally optimal control law. Notably, when $A=B=0$, a double-degeneration case takes place. In this case, the pitch angle can be selected as an arbitrary value belonging to $\left[-\frac{\pi}{2},\text{ }\frac{\pi}{2}\right]$, which may cause discontinuity in the acceleration shown in Eq. \eqref{eq1}. When integrating the trajectories, this case is examined, and this case does not take place in the simulation results presented in this Note. Subsequently, with the locally optimal control law, the concept of reflective-sail WSB structure is proposed.

\section{Reflective-Sail WSB Structure}\label{sec3}
In this section, we propose the concept of reflective-sail WSB structures. The configurations of reflective-sail WSB structures are then presented, providing a useful framework to analyze the dynamical behaviors around the Earth under the Sun-Earth PCR3BP with a reflective sail.
\subsection{Concept of Reflective-Sail WSB Structure}\label{sec3.1}
To construct reflective-sail escape trajectories, the reflective-sail WSB structure is proposed. The concept of WSB structure in the multi-body dynamics was firstly proposed by Belbruno and Miller \cite{belbruno1993sun}. Then, García and Gómez \cite{garcia2007note} further refined this concept. This concept has been widely used in the construction of low-energy transfer and ballistic capture trajectories \cite{belbruno1993sun,hyeraci2010method}. Also, Hyeraci and Topputo \cite{hyeraci2010method} implied the potential of the application of WSB structure in the construction of ballistic escape trajectories. The WSB structure denotes the initial periapsis set (also termed as a stable set) that can generate a stable motion. The initial periapsis satisfies $E_{20}<0$, where $E_2$ denotes the Keplerian energy with respect to the Earth and the subscript “0” denotes quantities corresponding to the initial epoch. Adopting the initial osculating eccentricity with respect to the Earth ($e_0$), $E_{20}$ can be further expressed as:
\begin{equation}
{E_{20}} = \frac{1}{2}\frac{\mu\left(e_0-1\right) }{r_{20}} \label{eq14}
\end{equation}
Therefore, $E_{20}<0$ is equivalent to $e_0<1$. With $e_0$, the initial periapsis state is expressed as (prograde case):
\begin{equation}
  {x_0} = {r_{20}}\cos {\theta _{\text{E0}}} + 1 - \mu  ,\text{ }{y_0} = {r_{20}}\sin {\theta _{\text{E0}}},\text{ }{u_0} =  - \left( {\sqrt {\frac{{\mu \left( {1 + e_0} \right)}}{{{r_{20}}}}}  - {r_{20}}} \right)\sin {\theta _{\text{E0}}},\text{ }{v_0} = \left( {\sqrt {\frac{{\mu \left( {1 + e_0} \right)}}{{{r_{20}}}}}  - {r_{20}}} \right)\cos {\theta _{\text{E0}}}
\label{eq15}
\end{equation}
where $l\left(\theta_\text{E}\right)$ denotes the phase angle with respect to the Earth in the Sun-Earth rotating frame (see Fig. \ref{fig_WSB}). Notably, the initial periapsis state for the retrograde case can also be obtained \cite{fu2026deep}. In this Note, we focus on the escape trajectories from the Earth. The prograde case is investigated for the practical mission requirement. With the aforementioned initial states presented in Eq. \eqref{eq15} ($e_0<1$), the stable/unstable motion can be defined. The trajectory departs the initial periapsis on $l\left(\theta_\text{E0}\right)$, and then the stable motion can be defined as the trajectory completes at least one revolution around the Earth and return $l\left(\theta_\text{E0}\right)$ with the intersection point satisfying $E_2\leq 0$ \cite{garcia2007note,topputo2009computation}, otherwise the trajectory has an unstable motion, as shown in Fig. \ref{fig_WSB}. Then, the mathematical definition of the WSB structure can be expressed as:

\begin{figure}[H]
\centerline{\includegraphics[width=0.4\textwidth]{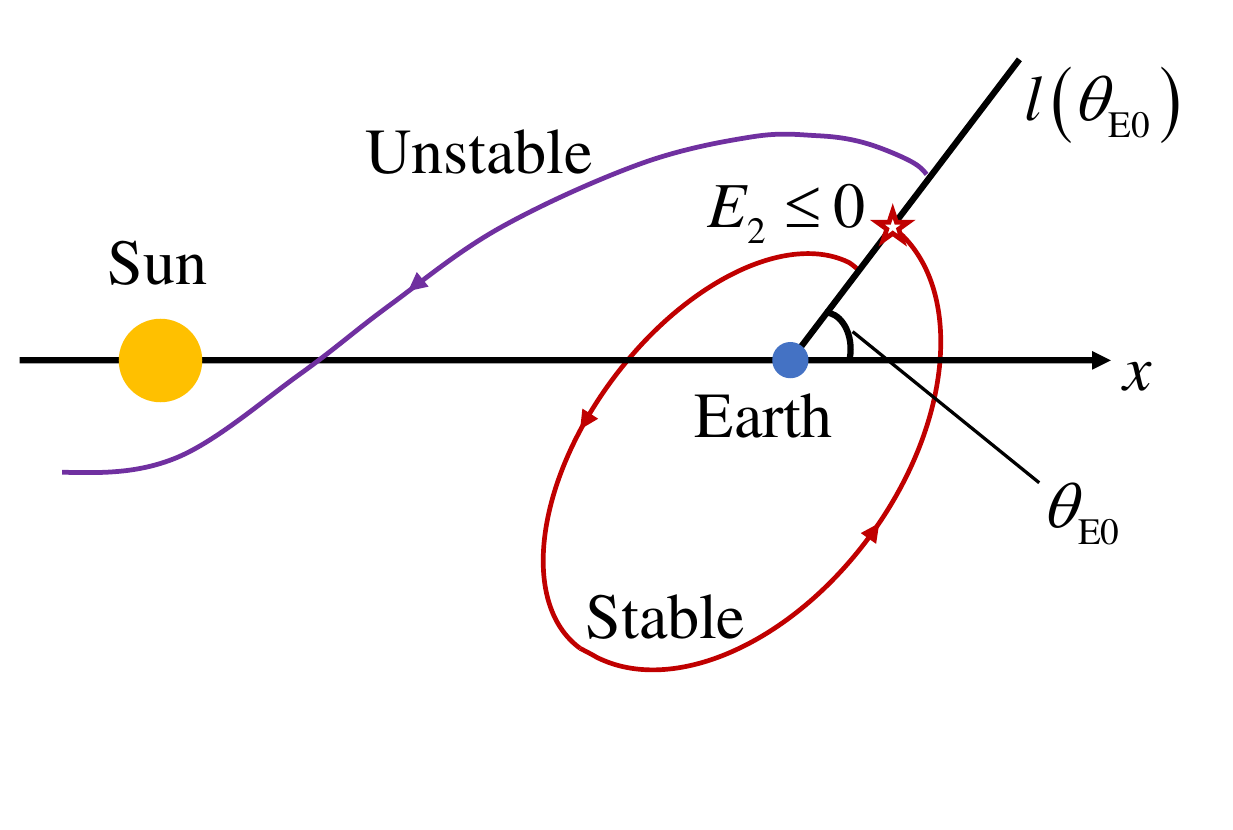}}
\caption{Schematic of the stable and unstable motions.}\label{fig_WSB}
\end{figure}

\begin{equation}
\mathcal{W}_1\left(e_0,\text{ }\beta,\text{ }\alpha_\text{LO}\right)=\left\{\left(x_0,\text{ }y_0\right)|E_2\leq 0 \text{ }\text{when}\text{ }\left|\theta_\text{E}-\theta_\text{E0}\right|=2\pi\right\}
\label{eq15_new}
\end{equation}
where the subscript “1” denotes the 1-stable set corresponding to at least one revolution. With a similar method, the $q$-stable set can be defined as:
\begin{equation}
{{\mathcal{W}_q}\left( e_0,\text{ }\beta,\text{ }\alpha_\text{LO} \right) = \left\{ {\begin{array}{*{20}{c}}
  {\left\{ {\left( {{x_0},{\text{ }}{y_0}} \right)|{E_2} \leq 0{\text{ when }}\left| {{\theta _{\text{E}}} - {\theta _{{\text{E0}}}}} \right| = 2q\pi } \right\}{\text{  }}\left( {q = 1} \right)} \\ 
  {\left\{ {\left( {{x_0},{\text{ }}{y_0}} \right)|\left( {{x_0},{\text{ }}{y_0}} \right) \in {\mathcal{W}_{q - 1}}\left( e_0,\text{ }\beta,\text{ }\alpha_\text{LO} \right){\text{ and }}{E_2} \leq 0{\text{ when }}\left| {{\theta _{\text{E}}} - {\theta _{{\text{E0}}}}} \right| = 2q\pi } \right\}{\text{  }}\left( {q > 1} \right)} 
\end{array}} \right.}
\end{equation}
The stable sets are termed as the forward stable sets if they are obtain by the forward-in-time integration, while they are termed as the backward stable sets if they are obtain by the backward-in-time integration (denoted as ${\mathcal{W}_{-q}}\left( e_0,\text{ }\beta,\text{ }\alpha_\text{LO} \right)$). The complementary of ${\mathcal{W}_{q}}\left( e_0,\text{ }\beta,\text{ }\alpha_\text{LO} \right)$ is termed as the $q$-unstable set, denoted as $\bar{\mathcal{W}}_{q}\left(e_0,\text{ }\beta,\text{ }\alpha_\text{LO}\right)$. In this Note, we focus on the configurations of $\mathcal{W}_1\left(e_0,\text{ }\beta,\text{ }\alpha_\text{LO}\right)$ because $\mathcal{W}_q\left(e_0,\text{ }\beta,\text{ }\alpha_\text{LO}\right) \in \mathcal{W}_1\left(e_0,\text{ }\beta,\text{ }\alpha_\text{LO}\right) \text{ }\left(q>1\right)$. Subsequently, the calculation of $\mathcal{W}_1\left(e_0,\text{ }\beta,\text{ }\alpha_\text{LO}\right)$ is performed.
\subsection{Calculation of Reflective-Sail WSB Structure}\label{sec3.2}
When calculating reflective-sail WSB structures, we select the sail’s lightness number $\beta$ up to 0.05 for the current practical consideration. Therefore, we set $\beta$ to $\beta=0.01,\text{ }0.03,\text{ and} \text{ }0.05$. For the construction of escape trajectories, we set $e_0$ to $e_0=0.5,\text{ }0.6,\text{ }0.7,\text{ }0.8,\text{ and}\text{ }0.9$. The position of initial periapses is set as follows: $r_{20} \in \left(0,\text{ }0.005\right]\text{ }\left(\text{LU}\right)$ with a step-size of $1\times 10^{-5}$ LU and $\theta_{\text{E}0} \in \left[0,\text{ }2\pi\right)$ with a step-size of $\pi/100$. The trajectory integration is performed by the dynamical equations shown in Eqs. \eqref{eq6} and \eqref{eq12}. The physical maximum integration time is set as ${T_{\max }} = \int_0^{10000} {{r_2}{\text{d}}s} $. The integration stops if one of the following criteria is satisfied:
\begin{enumerate}
\item The distance between the reflective sail and the Sun $r_1$ satisfies $r_1\leq 0.22\text{ LU}$ (for the safe distance \cite{wang2025optimal});
\item The position satisfies $x<-\mu$ and $y=0$.
\end{enumerate}
Then, the WSB structure ($\mathcal{W}_1\left(e_0,\text{ }\beta,\text{ }\alpha_\text{LO}\right)$) are recorded following the definition shown in Eq. \eqref{eq15_new}. To construct escape trajectories, the complementary of $\mathcal{W}_1\left(e_0,\text{ }\beta,\text{ }\alpha_\text{LO}\right)$ ($\bar{\mathcal{W}}_{1}\left(e_0,\text{ }\beta,\text{ }\alpha_\text{LO}\right)$) is also recorded. Subsequently, the configurations of reflective-sail WSB structures are presented, and their differences compared with those in the Sun-Earth PCR3BP are discussed.
\subsection{Configurations of Reflective-Sail WSB Structure}\label{sec3.3}
\begin{figure}[H]
\centerline{\includegraphics[width=0.9\textwidth]{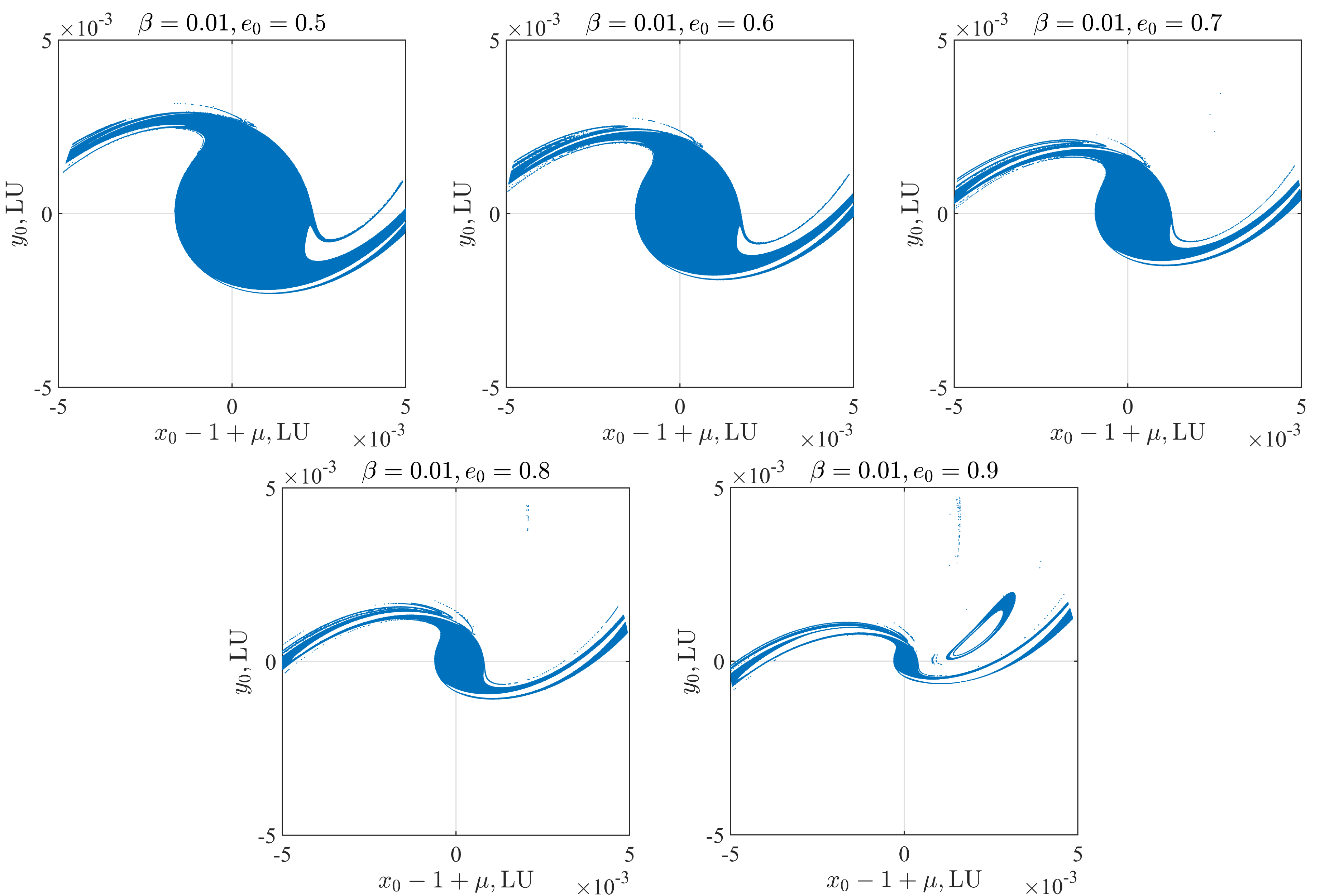}}
\caption{Configurations of reflective-sail WSB structures ($\beta=0.01$).}\label{fig_WSB_beta_0_01}
\end{figure}

\begin{figure}[H]
\centerline{\includegraphics[width=0.9\textwidth]{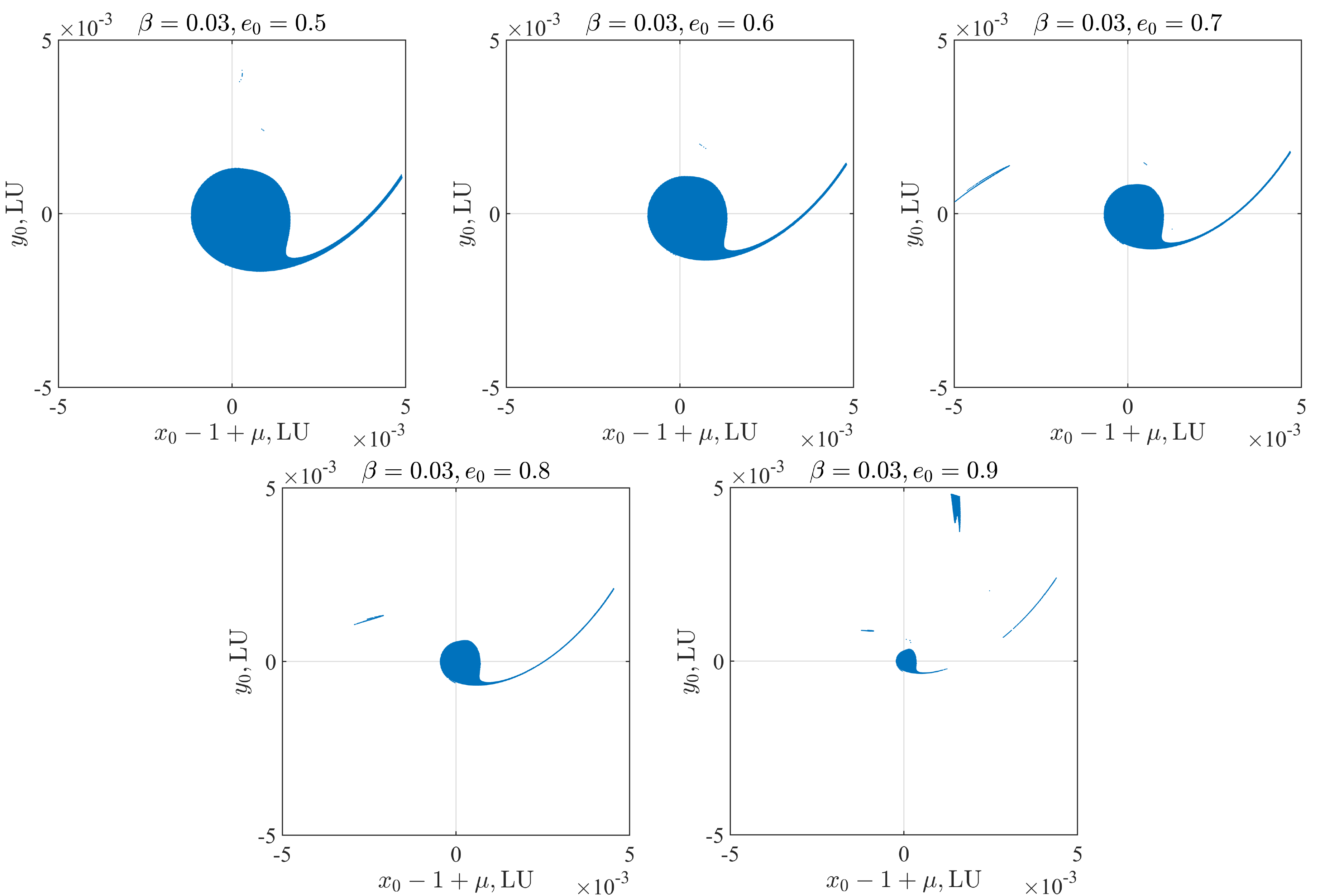}}
\caption{Configurations of reflective-sail WSB structures ($\beta=0.03$).}\label{fig_WSB_beta_0_03}
\end{figure}

\begin{figure}[H]
\centerline{\includegraphics[width=0.9\textwidth]{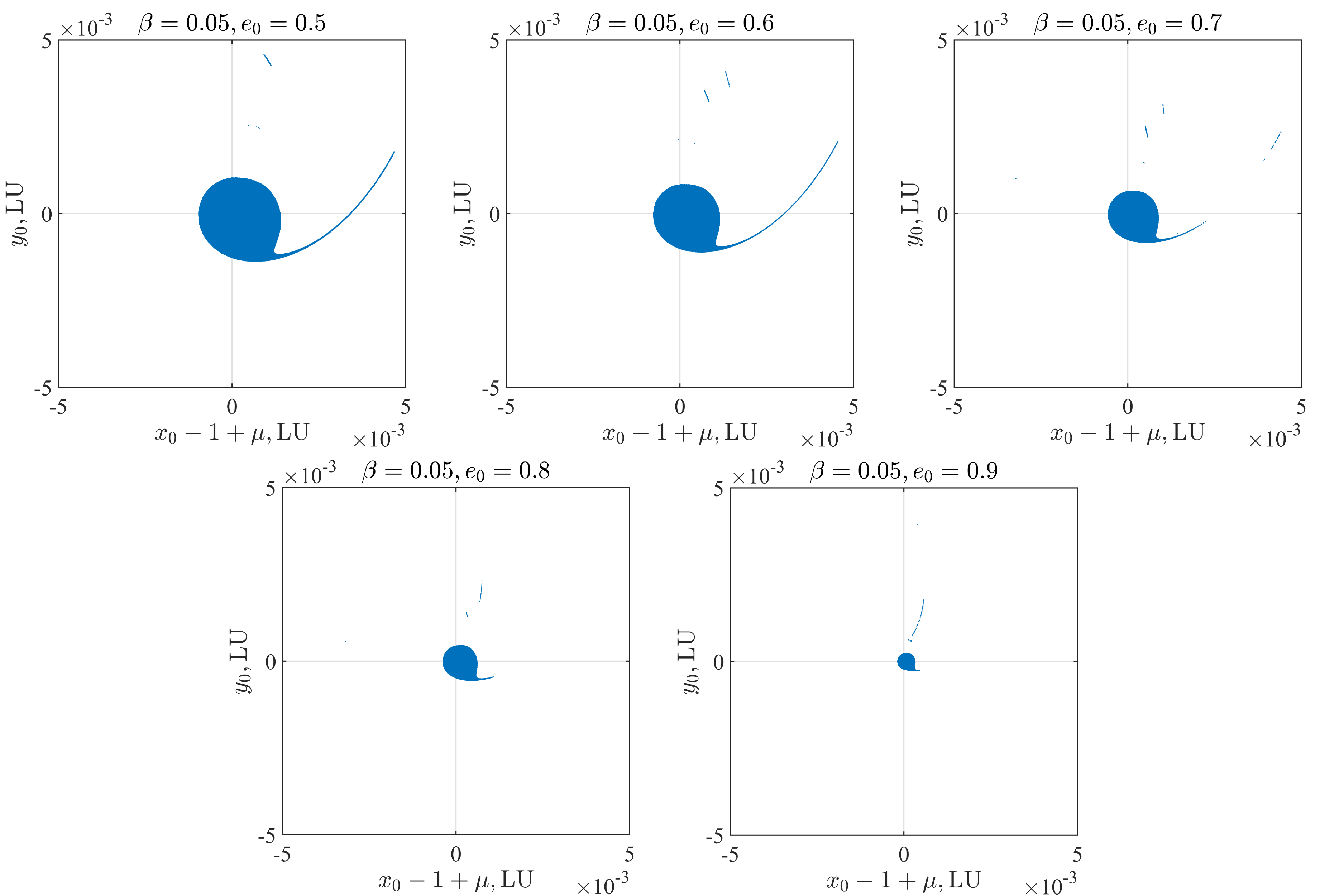}}
\caption{Configurations of reflective-sail WSB structures ($\beta=0.05$).}\label{fig_WSB_beta_0_05}
\end{figure}

\begin{figure}[H]
\centerline{\includegraphics[width=0.9\textwidth]{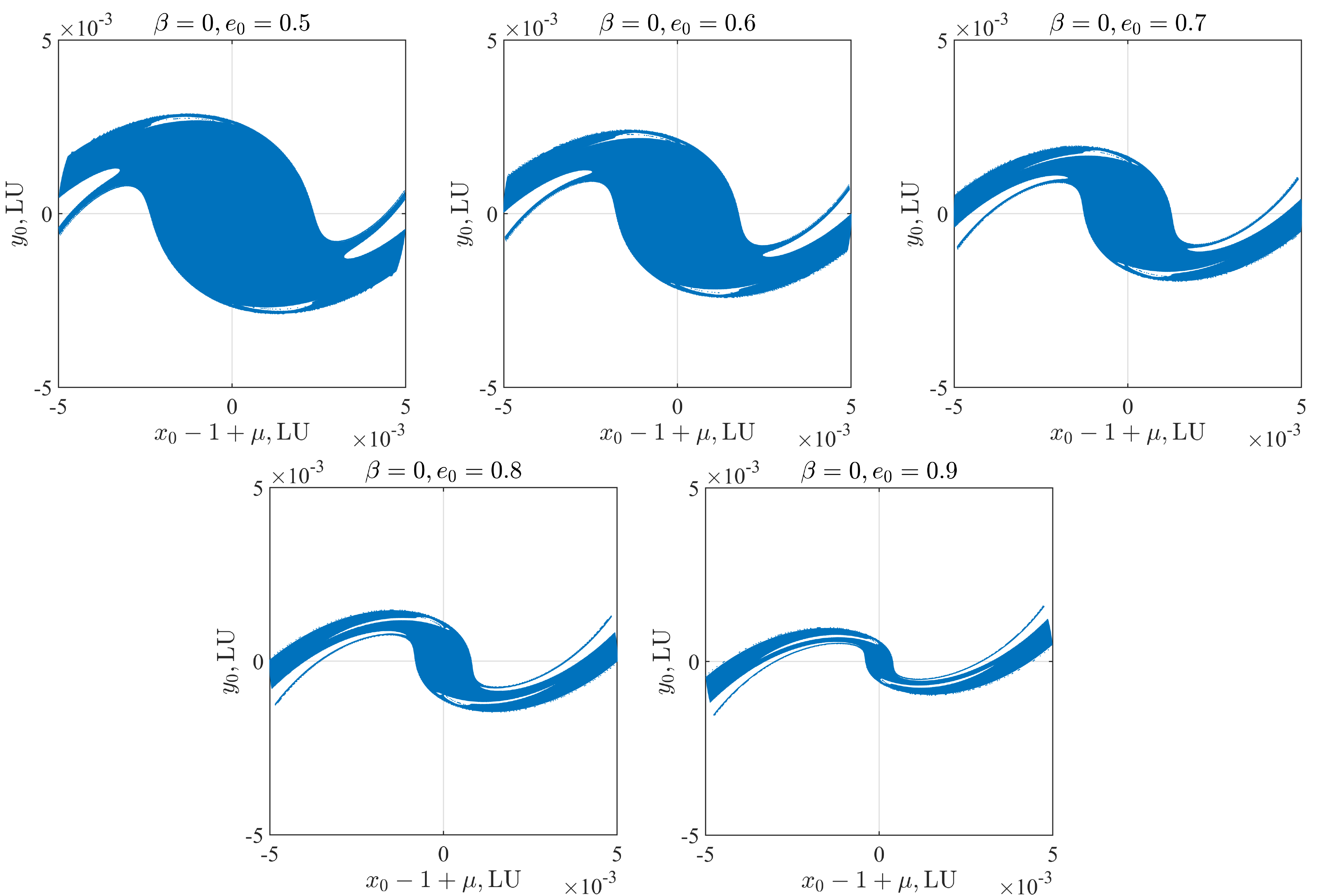}}
\caption{Configurations of WSB structures in the Sun-Earth PCR3BP.}\label{fig_WSB_PCR3BP}
\end{figure}

Figures \ref{fig_WSB_beta_0_01}-\ref{fig_WSB_beta_0_05} presents the configurations of reflective-sail WSB structures for $\beta$ from 0.01 to 0.05. To reveal the effects of the introduction of a reflective sail with the locally optimal control law, the comparison with the WSB structures in the Sun-Earth PCR3BP (i.e., $\beta=0$) is performed. The configurations of the WSB structures in the Sun-Earth PCR3BP are presented in Fig. \ref{fig_WSB_PCR3BP}. For $\beta=0.01$, it is observed that the configurations of reflective-sail WSB structures are similar to those in the Sun-Earth PCR3BP because of the relatively slight effects of the SRP acceleration. When $\beta=0.03$ and 0.05, the differences between the reflective-sail WSB structures and those in the Sun-Earth PCR3BP become more pronounced. It can be found that when the value of $\beta$ increases, the stable regions shrink toward the region around the Earth. The number of the stable points (i.e., initial periapses with the stable motions labeled by the blue scatters in Figs. \ref{fig_WSB_beta_0_01}-\ref{fig_WSB_PCR3BP}) for each case is summarized in Table \ref{tab:wsb_stable_points}. From this table, it can be observed that generally, the introduction of a reflective sail with the locally optimal control law facilitates the escape from the Earth, because the number of the stable points decreases as the value of $\beta$ increases. Meanwhile, as the value of $e_0$ increases, the number of the stable points decreases for each case, which is consistent with the conclusion in the Sun-Earth PCR3BP. However, when investigating the detailed distribution of the stable region, some regions where the use of a reflective sail does not facilitate the escape can also be observed. For convenience of description, the regions in Figs. \ref{fig_WSB_beta_0_01}-\ref{fig_WSB_PCR3BP} can be divided into four quadrants with the center of the Earth as the origin. Comparing Figs. \ref{fig_WSB_beta_0_01}-\ref{fig_WSB_beta_0_05} with Fig. \ref{fig_WSB_PCR3BP}, under the investigated ranges of $e_0$ and $\beta$, the regions corresponding to the unstable motion in the PCR3BP but the stable motion when a reflective sail with the locally optimal control law is used are typically located in the second quadrant. Based on the aforementioned discussion, the reflective-sail WSB structures provide a useful tool for quantitatively visualizing the effect of the introduction of a reflective sail with the locally optimal control law on the dynamics around the Earth, identifying the regions where the reflective sail can facilitate escape and those where it cannot. Subsequently, to quantitatively evaluate the improved performance of introducing a reflective sail, escape trajectories are constructed, and escape performance is analyzed accordingly. 

\begin{table}[htb!]
\centering
\renewcommand{\arraystretch}{1.3}
\caption{Number of the stable points for different WSB structures under different sail lightness numbers and initial eccentricities.}
\label{tab:wsb_stable_points}
\begin{tabular}{c|c|c|c}
\hline
\multicolumn{1}{c}{Type of WSB Structures} &
\multicolumn{1}{c}{$\beta$} &
\multicolumn{1}{c}{$e_0$} &
\multicolumn{1}{c}{Number of the Stable Points} \\
\hline

\multirow{15}{*}{Reflective Sail}
& \multirow{5}{*}{$0.01$} & 0.5 & 455739 \\
& & 0.6 & 367235 \\
& & 0.7 & 278900 \\
& & 0.8 & 189937 \\
& & 0.9 & 105593 \\
\cline{2-4}

& \multirow{5}{*}{$0.03$} & 0.5 & 294659 \\
& & 0.6 & 238038 \\
& & 0.7 & 180731 \\
& & 0.8 & 121728 \\
& & 0.9 & 62008 \\
\cline{2-4}

& \multirow{5}{*}{$0.05$} & 0.5 & 237878 \\
& & 0.6 & 191923 \\
& & 0.7 & 145169 \\
& & 0.8 & 97663 \\
& & 0.9 & 48802 \\
\hline

\multirow{5}{*}{PCR3BP}
& \multirow{5}{*}{--} & 0.5 & 583388 \\
& & 0.6 & 470446 \\
& & 0.7 & 362431 \\
& & 0.8 & 257148 \\
& & 0.9 & 152901 \\
\hline

\end{tabular}
\end{table}

\section{Construction and Evaluation of Escape Trajectories Using Reflective-Sail WSB Structures}\label{sec4}
This section presents the construction method of escape trajectories based on the reflective-sail WSB structures. Then, escape performance of the obtained solutions is evaluated in terms of time of flight (TOF) and estimated hyperbolic excess velocity ($v_{\infty}$). Comparison with ballistic escape trajectories is also performed. Results show that Pareto fronts of the $\left(\text{TOF},\text{ }v_{\infty}\right)$ distributions of the obtained solutions using a reflective sail are trended to be distributed in the upper-left corner of those of ballistic escape trajectories in the Sun-Earth PCR3BP, indicating improved escape performance characterized by a balance between shorter TOF and higher $v_{\infty}$.
\subsection{Construction of Escape Trajectories}\label{sec4.1}
\begin{figure}[H]
\centerline{\includegraphics[width=0.6\textwidth]{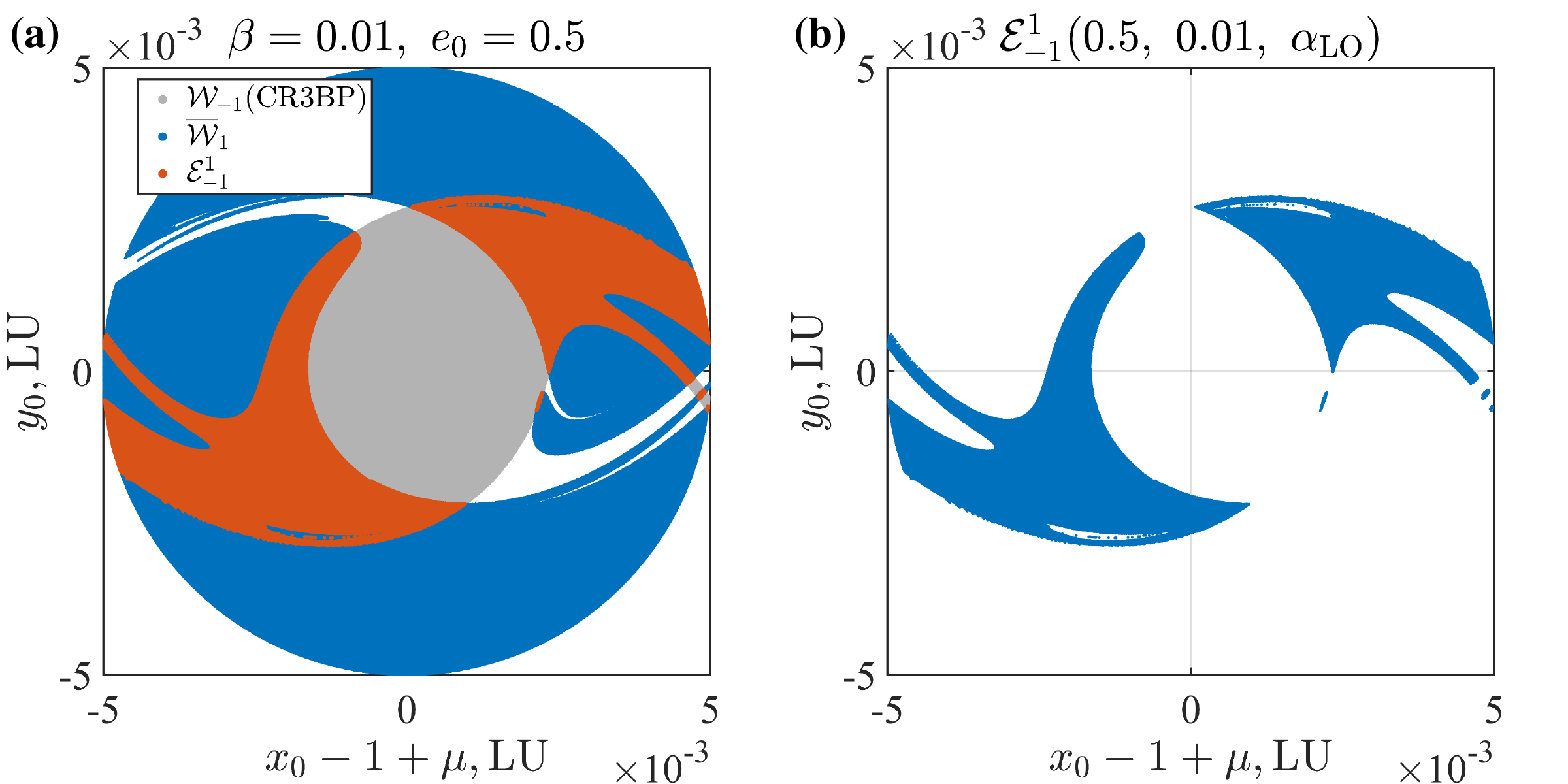}}
\caption{An example of the intersection shown in Eq. \eqref{eq24} and the escape set. (a) Intersection of the stable/unstable sets. (b) Escape set.}\label{fig_intersection_sample}
\end{figure}

\begin{figure}[H]
\centerline{\includegraphics[width=0.4\textwidth]{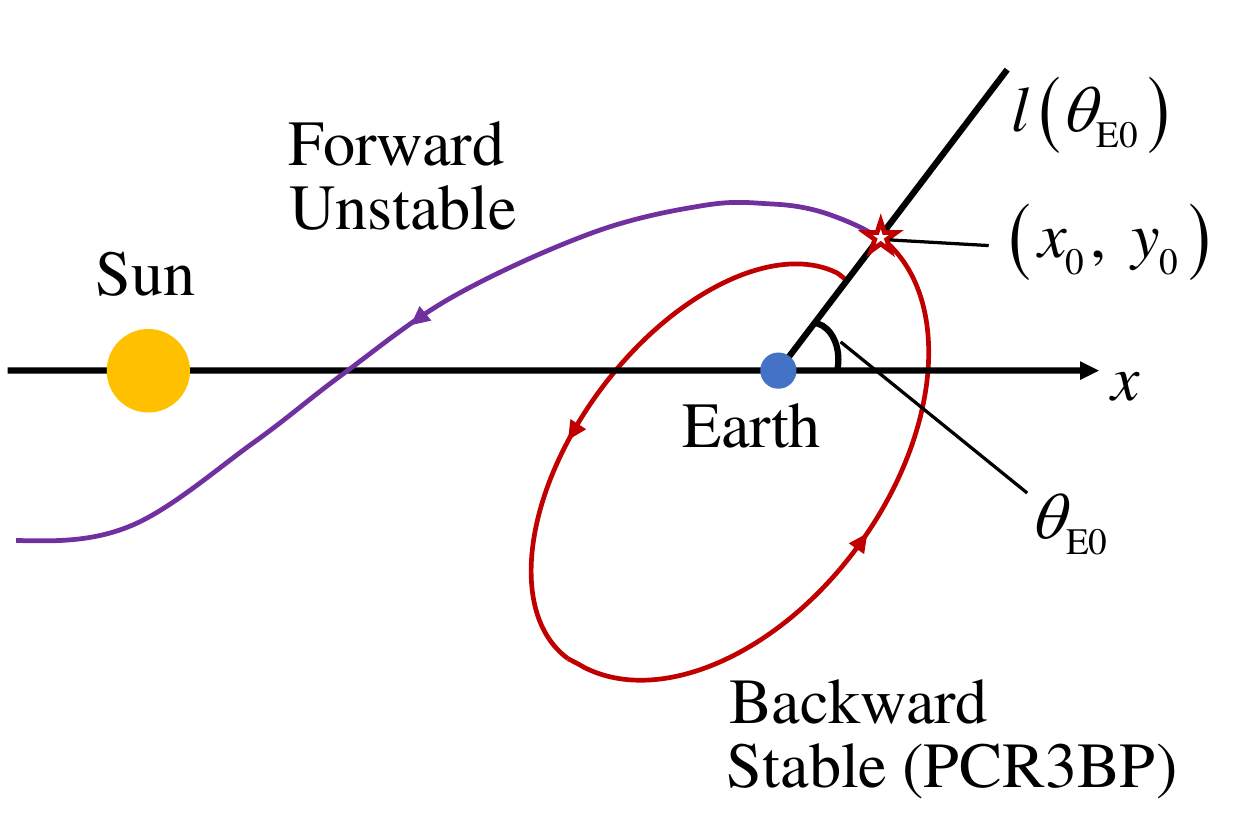}}
\caption{Schematic of the escape trajectory generated by the escape set.}\label{escape_trajectory}
\end{figure}

In this Note, we use the method proposed by Hyeraci and Topputo \cite{hyeraci2010method} to construct escape trajectories. To accommodate the scenarios considered in this Note (i.e., reflective-sail escape trajectories), we make several modifications to the method. The core of the construction method proposed by Hyeraci and Topputo is the construction of the ballistic capture/escape sets, which are obtained by the intersection of the stable/unstable sets in the PCR3BP. In this Note, however, we consider the construction of escape sets by intersecting the forward unstable sets under the Sun-Earth PCR3BP with a reflective sail with the backward stable sets under the Sun-Earth PCR3BP. The justification of this setting is discussed in the following texts. In this Note, the escape sets in this Note are constructed as follows:
\begin{equation}\label{eq24}
\mathcal{E}_{-1}^1\left(e,\text{ }\beta,\text{ }\alpha_\text{LO}\right)={\mathcal{W}}_{-1}\left(e_0,\text{ }0,\text{ --}\right) \cap \bar{\mathcal{W}}_1\left(e_0,\text{ }\beta,\text{ }\alpha_\text{LO}\right)
\end{equation}
where ${\mathcal{W}}_{-1}\left(e_0,\text{ }0,\text{ --}\right)$ denotes the backward stable sets in the Sun-Earth PCR3BP. According to the symmetric properties of the PCR3BP:
\begin{equation}
{\left( {x,{\text{ }}y,{\text{ }}u,{\text{ }}v,{\text{ }}t} \right) \to \left( {x,{\text{ }} - y,{\text{ }} - u,{\text{ }}v,{\text{ }} - t} \right)} \label{eq_sys}
\end{equation}
we obtain the backward stable sets by transforming $\left(x_0,\text{ }y_0\right)$ belonging to ${\mathcal{W}}_{-1}\left(e_0,\text{ }0,\text{ --}\right)$ to $\left(x_0,\text{ }-y_0\right)$. An example of the intersection shown in Eq. \eqref{eq24} and the corresponding escape set is presented in Fig. \ref{fig_intersection_sample}. 
Following Eq. \eqref{eq24}, we totally obtain 20 escape sets, including ballistic escape sets in the Sun-Earth PCR3BP for comparison. The periapses in the escape sets denote these initial states which can generate trajectories with the unstable motion by forward-in-time integration, while generating trajectories with the stable motion by back-in-time integration in the Sun-Earth PCR3BP, as shown in Fig. \ref{escape_trajectory}. The physical meaning of this construction is as follows: the reflective sail is initially launched with the sail undeployed and coasts under the Sun-Earth gravity. When it passes through a periapsis, the sail is deployed, and the locally optimal control law is performed. The constraint of the backward stable sets in the Sun-Earth PCR3BP ensures that the spacecraft is not moving in a backward escape trajectory before the deployment of the sail. In the simulations, for initial states in the escape sets, we perform backward-in-time integration to obtain trajectories in the Sun-Earth PCR3BP. In particular, we select the segment satisfying $\left| {{\theta _{\text{E}}} - {\theta _{{\text{E}}0}}} \right| \leqslant 2\pi $ as the backward segment of the escape trajectories. The forward-in-time integration is performed to obtain the forward segment of the escape trajectories. The integration stops when the trajectories satisfy conditions (1) or (2) mentioned in Section \ref{sec3.2}. We select the trajectory segment from the initial states in the escape sets to the final epoch at which it intersects a disk around the Earth, i.e., where $r_2=R_d$, as the forward segment of the escape trajectory. Notably, if the type of forward instability is $E_2>0$ when $\left| {{\theta _{\text{E}}} - {\theta _{{\text{E}}0}}} \right| = 2\pi $, the integration is continued until the trajectories satisfy conditions (1) or (2) mentioned in Section \ref{sec3.2} after the reflective sail returns $l\left({\theta _{\text{E}}}\right)$. The maximum integration time of this continuation is set as ${T_{\max \text{con}}} = \int_0^{300000} {{r_2}{\text{d}}s} $. Trajectories impacting the Earth (i.e., trajectories existing the states satisfying $r_2=R_\text{E}$) are excluded. Then, to quantitatively evaluate the escape performance improved by the introduction of SRP acceleration, parameters revealing escape performance are presented, including TOF and estimated hyperbolic excess velocity. 
The TOF is defined as the sum of backward time and forward time:
\begin{equation}
{\text{TOF}} = {t_{{\text{backward}}}} + {t_{{\text{forward}}}} \label{eq26}
\end{equation}
where ${t_{{\text{backward}}}}$ records the time of flight of the backward segment, while ${t_{{\text{forward}}}}$ records the time of flight of the forward segment. A shorter TOF means a more efficient escape. The hyperbolic excess velocity is estimated by the value of the velocity with respect to the Earth when the reflective sail satisfies $r_2=R_d$ for the final time:
\begin{equation}
{v_\infty } \approx \sqrt {{{\left( {{u_f} - {y_f}} \right)}^2} + {{\left( {{v_f} + {x_f} + \mu  - 1} \right)}^2}} \label{eq27}
\end{equation}
where the subscript “\textit{f}” denotes quantities corresponding to the epoch when trajectory states satisfy $r_2=R_d$ for the final time. A higher ${v_\infty }$ means a better reachability and a shorter time for interplanetary transfers. Subsequently, we present an evaluation of the improvement of the introduction of a reflective sail with the locally optimal control law. To perform a sensitivity analysis, we set $R_d$ as 0.01, 0.02, and 0.03 LU, and present the corresponding comparison between reflective-sail escape trajectories and ballistic escape trajectories in the Sun-Earth PCR3BP.
\subsection{Evaluation of Escape Trajectories}\label{sec4.2}
\begin{figure}[H]
\centerline{\includegraphics[width=0.9\textwidth]{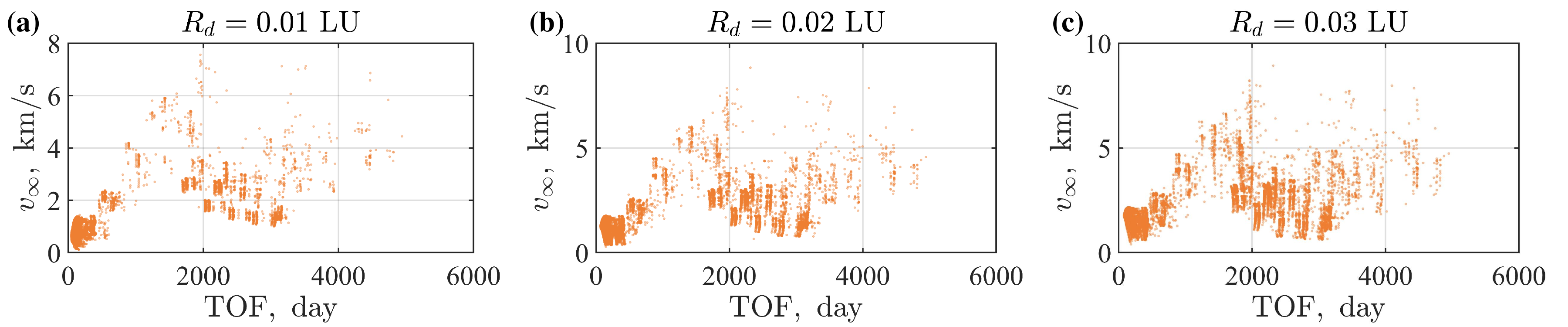}}
\caption{The $\left(\text{TOF},\text{ }v_{\infty}\right)$ map under $\beta=0.05$ and $e_0=0.9$. (a) $R_d=0.01\text{ LU}$; (b) $R_d=0.02\text{ LU}$; (c) $R_d=0.03\text{ LU}$.}\label{fig_solution_space}
\end{figure}

Figure \ref{fig_solution_space} presents the $\left(\text{TOF},\text{ }v_{\infty}\right)$ map of all obtained escape trajectories constructed by escape sets under $\beta=0.05$ and $e_0=0.9$ for $R_d=0.01,\text{ }0.02,\text{ and}\text{ }0.03\text{ LU}$. From this figure, it can be found that the value of $R_d$ can affect the distribution of solutions due to the introduction of the locally optimal control law. The median and 90th percentile of the TOF distributions of these three cases are summarized in Table \ref{50_p90}. As most of solutions considered as escape trajectories with relatively short TOF, some solutions with very long TOF can also be observed. For practical consideration, we select solutions with $\text{TOF}<200\text{ days}$, and evaluate the escape performance in terms of TOF and estimated $v_{\infty}$. The analysis is performed for all the solutions constructed by the aforementioned 20 escape sets.

\begin{figure}[H]
\centerline{\includegraphics[width=0.9\textwidth]{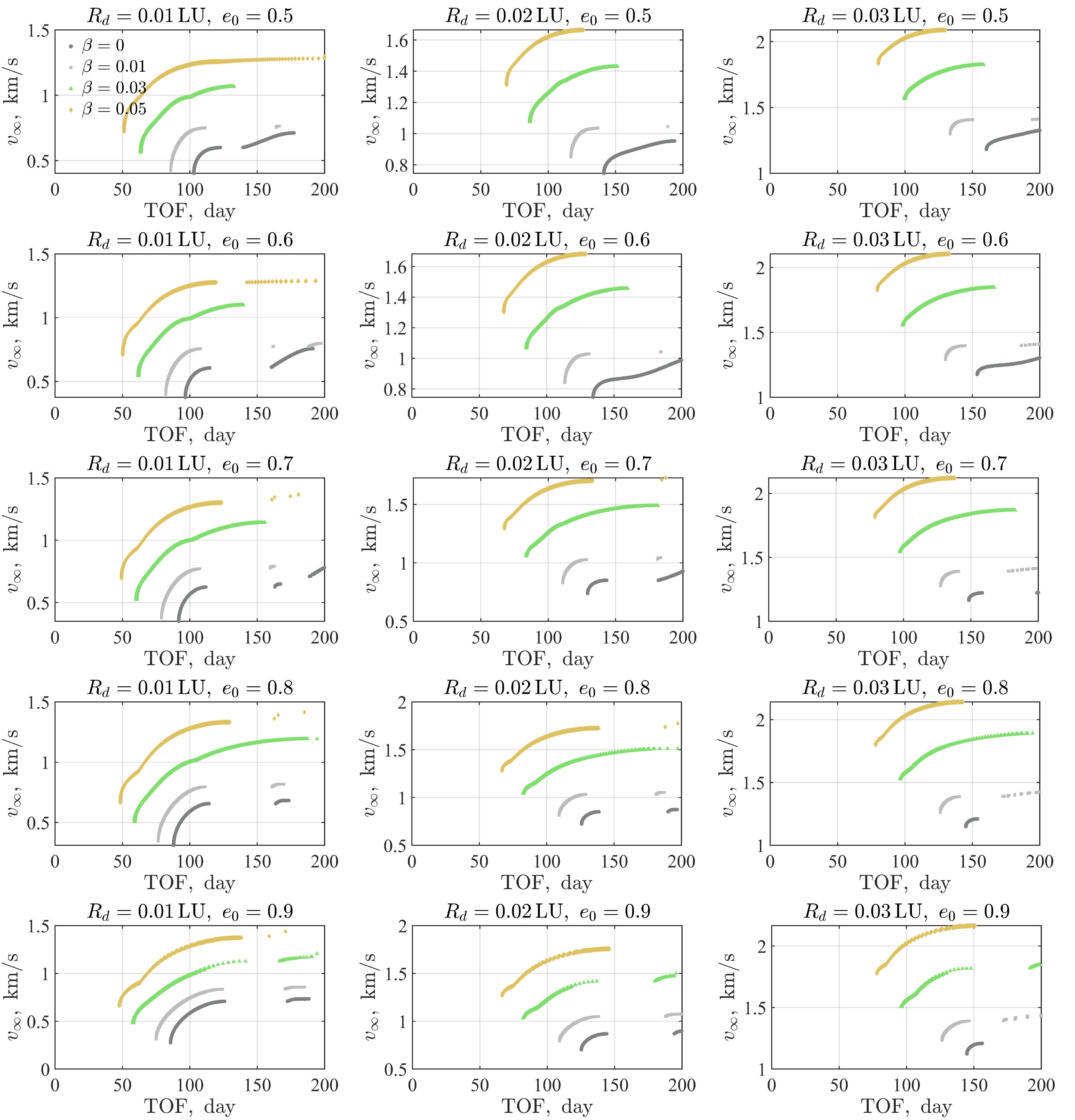}}
\caption{Pareto fronts of obtained solutions within 200 days.}\label{pareto_all}
\end{figure}

\begin{table}[!htb]
\caption{Median and 90th percentile of the TOF distributions}\label{50_p90}%
\centering
\renewcommand{\arraystretch}{1.5}
\begin{tabular}{@{}lll@{}}
\hline
$R_d,\text{ LU}$ & Median, Day  & 90th Percentile, Day\\
\hline
0.01    & 137   & 189    \\
0.02    & 161   & 292  \\
0.03    & 181   & 580  \\
\hline
\end{tabular}
\end{table}
For all obtained solutions within 200 days constructed by 20 escape sets for $R_d=0.01,\text{ }0.02,\text{ and}\text{ }0.03\text{ LU}$, we construct the $\left(\text{TOF},\text{ }v_{\infty}\right)$ map, and extract their Pareto fronts \cite{topputo2013optimal,oshima2019low} considering shorter TOF and higher $v_{\infty}$, as shown in Fig. \ref{pareto_all}. It can be found that although different values of $R_d$ cause different $\left(\text{TOF},\text{ }v_{\infty}\right)$ distributions and Pareto fronts, the qualitative trends in the escape performance improved by the introduction of a reflective sail with the locally optimal control law remain unchanged. From Fig. \ref{pareto_all}, when the value of $\beta$, Pareto fronts trend to be distributed in the upper-left corner of those with lower values of $\beta$, including $\beta=0$, i.e., the Sun-Earth PCR3BP. This comparison clearly reveals the advantages of introducing a reflective sail with the locally optimal control law in the improvement of escape performance. Moreover, the increase of the value of $\beta$ brings a continuous improvement of escape performance evaluated by TOF and estimated $v_{\infty}$. For this reason, we select $\beta=0.05$ for the analysis of the typical escape trajectory subsequently. 
\subsection{Analysis of Typical Escape Trajectory}\label{sec4.3}
\begin{figure}[H]
\centerline{\includegraphics[width=0.3\textwidth]{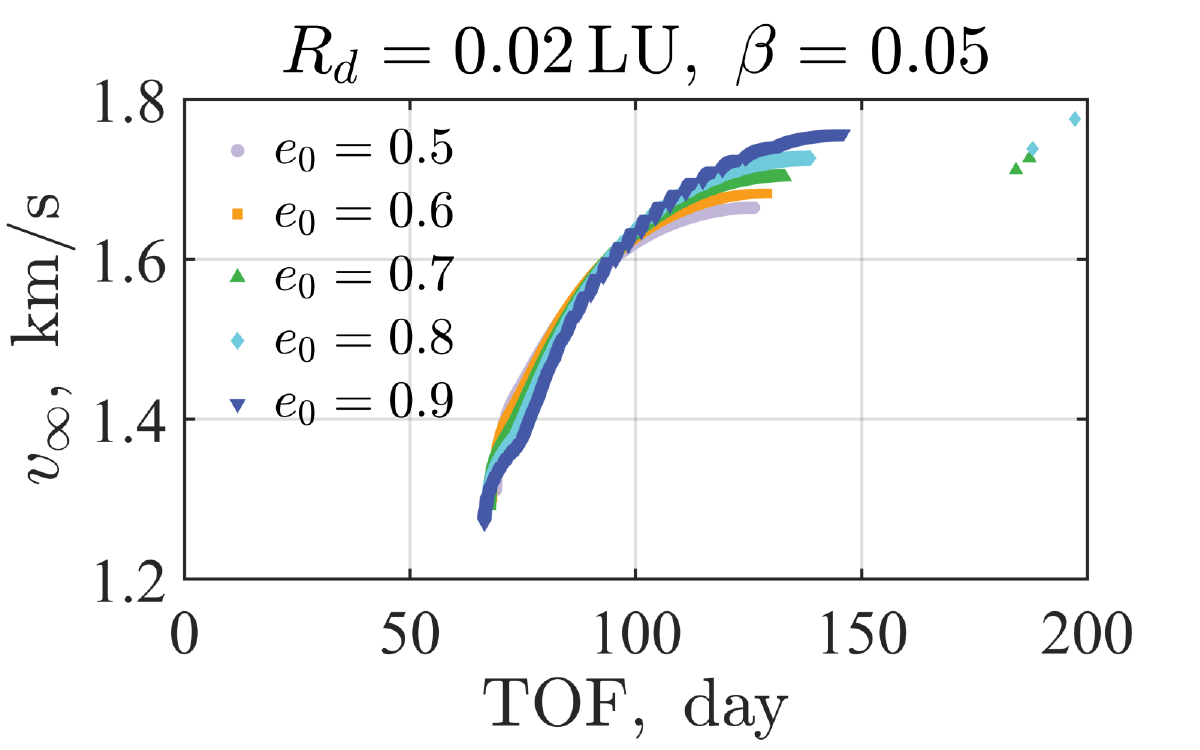}}
\caption{Pareto fronts under $\beta=0.05$ for all investigated values of $e_0$.}\label{pareto_0_05}
\end{figure}

Due to the similarity of the qualitative trends of Pareto fronts for the investigated values of $R_d$, we select $R_d=0.02\text{ LU}$ to perform the subsequent analysis. Figure \ref{pareto_0_05} presents Pareto fronts under $\beta=0.05$ for all investigated values of $e_0$. It can be observed that considering the samples on Pareto fronts, the increase of the value of $e_0$ generally causes the increase of the maximum $v_{\infty}$ of the obtained solutions within 150 days (notably, some higher $v_{\infty}$ solutions with lower values of $e_0$ can also be found, at the cost of longer TOF). Therefore, we select the maximum $v_{\infty}$ solution under $e_0=0.9$ to present a typical escape trajectory.

\begin{figure}[H]
\centerline{\includegraphics[width=0.9\textwidth]{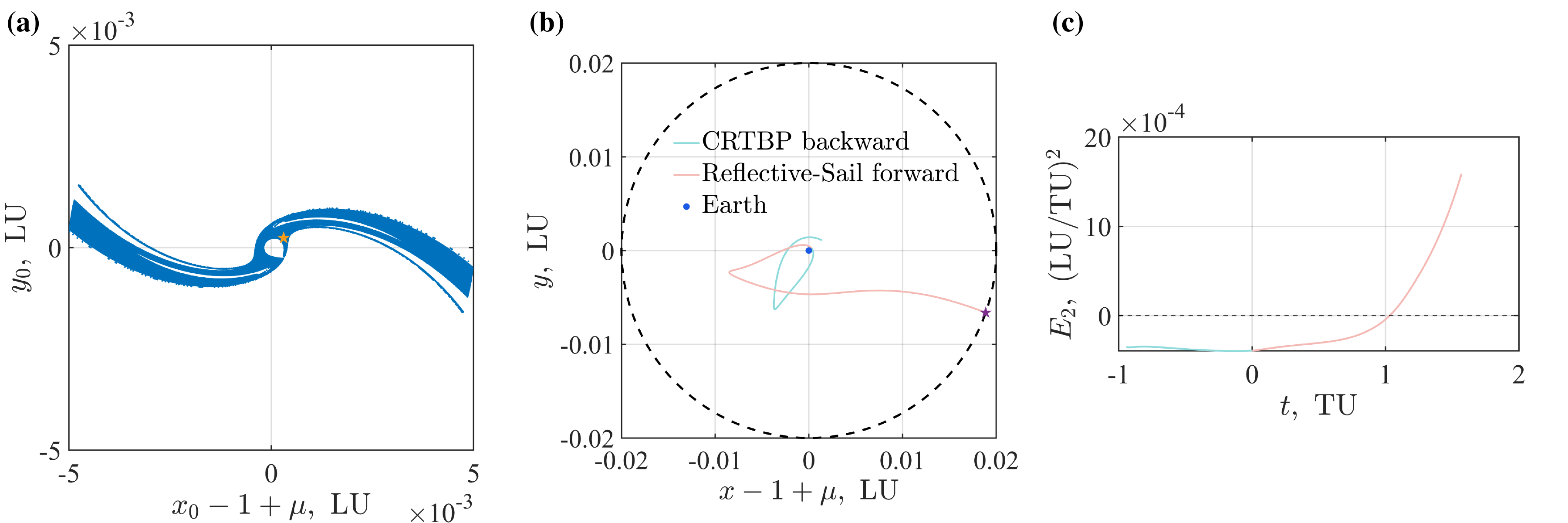}}
\caption{Maximum $v_{\infty}$ solution under $e_0=0.9$. (a) Initial states in the escape set; (b) Escape trajectory; (c) Time history of $E_2$.}\label{escape_trajectory_sample}
\end{figure}

\begin{figure}[H]
\centerline{\includegraphics[width=0.6\textwidth]{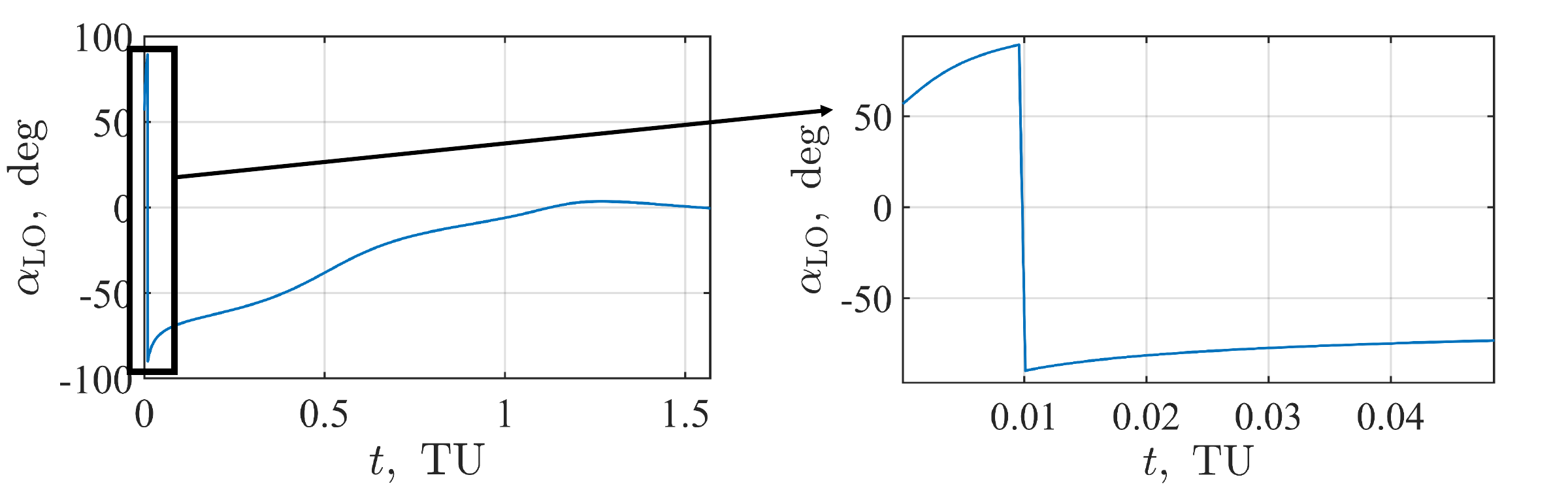}}
\caption{Time history of $\alpha_\text{LO}$.}\label{LO}
\end{figure}

Figure \ref{escape_trajectory_sample} presents the the maximum $v_{\infty}$ solution under $e_0=0.9$. The orange star in Fig. \ref{escape_trajectory_sample} (a) denotes the initial state in the escape set. Figure \ref{escape_trajectory_sample} (b) presents the escape trajectory, and Fig. \ref{escape_trajectory_sample} (c) presents the time history of $E_2$. The reflective sail firstly coasts under the Sun-Earth gravity, and deploys the sail to perform the locally optimal control law (the time history of $\alpha_\text{LO}$ can be found in Fig. \ref{LO}) to escape from the Earth. Accordingly, the value of $E_2$ changes from negative to positive. As shown in Fig. \ref{LO}, $\alpha_\text{LO}$ changes from 90 deg to $-90$ deg for one time, and the time history of $\alpha_\text{LO}$ can be considered consistent with the results presented by McInnes \cite{mcinnes2004solar}. The $\left(\text{TOF},\text{ }v_{\infty}\right)$ of this solution is 146 days and 1.7573 km/s. The aforementioned discussion is performed under the Sun-Earth PCR3BP with a reflective sail. To further consider the practical application, the shadow and higher-fidelity dynamical models \cite{belbruno1993sun,macdonald2005realistic,zambrano2026solar} can be considered in the calculation of reflective-sail WSB structures and escape trajectory construction.

\section{Conclusion}\label{sec5}
This Note proposes the concept of reflective-sail weak stability boundary structures and accordingly constructs and analyzes escape trajectories from the Earth in the context of the Sun-Earth planar circular restricted three-body problem with a reflective sail. To facilitate escape, the locally optimal control law to maximize the time derivative of the Keplerian energy with respect to the Earth is adopted. Levi-Civita regularization about the Earth is derived to address the singularity caused by the Earth, and the configurations of reflective-sail weak stability boundary structures are calculated to provide initial states for constructing escape trajectories and information about regions where escape is facilitated compared with the purely Sun-Earth planar circular restricted three-body problem. A construction method of escape trajectories is presented based on the reflective-sail weak stability boundary structures. The escape performance, including time of flight and estimated hyperbolic excess velocity, is analyzed. Comparison with ballistic escape trajectories in the Sun-Earth PCR3BP is also performed, indicating improved escape performance characterized by shorter time of flight and higher hyperbolic excess velocity. This Note performs a systematic analysis of escape trajectories using a reflective sail with the locally optimal control law, providing a useful insight into the reflective-sail multi-body dynamics and the construction of reflective-sail escape trajectories.

\section*{Acknowledgments}
This work was supported by the National Natural Science Foundation of China (Grant Nos. 12372044, 12525204, and U23B6002).

\bibliography{sample}

\end{document}